\documentclass[usenatbib]{mn2e}
\usepackage{amssymb}
\usepackage{graphicx}
\usepackage{color}

\newcommand{\be}{\begin{equation}}
\newcommand{\ee}{\end{equation}}
\newcommand{\vecB}{\bmath B}
\newcommand{\vecA}{\bmath A}
\newcommand{\bdiff}{\eta}
\newcommand{\econd}{\sigma}
\newcommand{\econdei}{\sigma_\mathrm{ei}}
\newcommand{\econdph}{\sigma_\mathrm{ph}}
\newcommand{\tohm}{\tau_\mathrm{Ohm}}
\newcommand{\Tc}{T_\mathrm{c}}
\newcommand{\Ts}{T_\mathrm{s}}
\newcommand{\gs}{g_\mathrm{s}}
\newcommand{\kb}{k_\mathrm{B}}
\newcommand{\sigsb}{\sigma_\mathrm{SB}}
\newcommand{\sigth}{\sigma_\mathrm{T}}
\newcommand{\tdiff}{K}

\newcommand{\opact}{\kappa}
\newcommand{\opacc}{\kappa_\mathrm{c}}
\newcommand{\opacr}{\kappa_\mathrm{r}}
\newcommand{\opacff}{\kappa_\mathrm{ff}}
\newcommand{\opaces}{\kappa_\mathrm{es}}
\newcommand{\Pdot}{\dot{P}}
\newcommand{\Edot}{\dot{E}}
\newcommand{\Bp}{B_\mathrm{p}}
\newcommand{\Bdot}{\dot{B}_\mathrm{p}}
\newcommand{\denssub}{\rho_\mathrm{sub}}
\newcommand{\rsub}{r_\mathrm{sub}}
\newcommand{\rsubwidth}{\delta_r}
\newcommand{\Bstar}{B_\ast}

\title[Magnetic field evolution in CCOs]{
Evolution of a buried magnetic field in the central compact object
neutron stars}
\author[W.~C.~G. Ho]{Wynn C.~G. Ho$^1$\thanks{email: wynnho@slac.stanford.edu}\\
$^1$School of Mathematics, University of Southampton, Southampton, SO17 1BJ}

\begin{document}
\pagerange{\pageref{firstpage}--\pageref{lastpage}} \pubyear{2011}
 
\maketitle

\label{firstpage}

\begin{abstract}
The central compact objects are a newly-emerging class of young
neutron stars near the centre of supernova remnants.
From X-ray timing and spectral measurements, their magnetic
fields are determined to be $\sim 10^{10}-10^{11}$~G, which is
significantly lower than that found on most pulsars.
Using the latest electrical and thermal conductivity calculations, we
solve the induction equation to determine the evolution of a buried
crustal or core magnetic field.
We apply this model of a buried field to explain the youth and low
observed magnetic field of the central compact objects.
We obtain constraints on their birth magnetic field
and depth of submergence (or accreted mass).
Measurement of a change in the observed magnetic field strength would
discriminate between the crustal and core fields and could yield
uniquely the birth magnetic field and submergence depth.
If we consider the central compact objects as a single neutron star
viewed at different epochs, then we constrain the magnetic field at
birth to be $\sim (6-9)\times 10^{11}$~G.
A buried magnetic field can also explain their location in an
underpopulated region of the spin period-period derivative plane
for pulsars.
\end{abstract}


\section{Introduction} \label{sec:intro}

The magnetic fields on the surface of neutron stars (NSs) are known to span
a large range:
from $B\sim 10^8-10^9$~G for millisecond pulsars and NSs in low-mass X-ray
binaries, through $10^{12}-10^{13}$~G for normal radio pulsars,
to $10^{14}-10^{15}$~G for magnetars.
The primary method for determining the magnetic field is by measuring
the spin period $P$ and spin period derivative $\Pdot$.
Then, by assuming that the pulsar rotational energy is lost through
magnetic dipole radiation, one obtains the relation for the magnetic
field at the pole $\Bp$,
\be
P\Pdot = \frac{\gamma}{2}\Bp^2
\quad \Leftrightarrow \quad
\Bp = 6.4\times 10^{19}\mbox{ G }(P\Pdot)^{1/2},
\label{eq:pdot}
\ee
where $\gamma=4\pi^2R^6\sin^2\alpha/3c^3I
 = 4.884\times 10^{-40}\mbox{ s G$^{-2}$ }R_6^6I_{45}^{-1}\sin^2\alpha$,
$R$ and $I$ are the NS radius and moment of inertia, respectively,
$\alpha$ is the angle between the NS rotation and magnetic axes,
and $R_6=R/10^6\mbox{ cm}$ and $I_{45}=I/10^{45}\mbox{ g cm$^2$}$
(\citealt{gunnostriker69}; see also \citealt{shapiroteukolsky83,spitkovsky06}).
Note that a coefficient of 3.2 is often used in the literature, so that
the magnetic field that is inferred is the equatorial value;
since we are modelling the field evolution at the pole, we will hereafter
only refer to the magnetic field at the pole.

The evolution of NS magnetic fields is uncertain.
Radio pulsars are seen to span a large range of ages
($\sim 10^3-10^9$~yr), magnetars are relatively young ($\sim 10^4$~yr), and
millisecond pulsars are old ($\sim 10^8-10^9$~yr).
The diffusion and decay of the magnetic field occurs on the Ohmic
timescale
\be
\tohm = 4\pi\econd L^2/c^2 \sim 4\times 10^5\mbox{ yr }
 \left(\econd/10^{24}\mbox{ s$^{-1}$}\right)\left(L/1\mbox{ km}\right)^2,
 \label{eq:tohm}
\ee
where $\econd$ is the electrical conductivity
and $L$ is the lengthscale over which the decay occurs;
here we have taken $L$ to be approximately the size of the NS crust
(see Fig.~\ref{fig:acc}).
Note however that there is a large range in $\tohm$ due to the density
dependence of the conductivity and the large density gradient in the crust.
Thus NSs born with $B\gtrsim 10^{12}$~G survive for long times at these
field strengths.
A manifestation of these high fields is the observed pulsar/magnetosphere
activity seen in young (and old) NSs, such as in the Crab pulsar which is
only $\approx 10^3$~yr old.
The low magnetic fields which exist on millisecond pulsars (and on the NS
primaries in low-mass X-ray binaries) are thought to be due to burial of
the field by mass accretion from a companion star.
However, several young ($\lesssim 10^4$~yr) NSs are believed to have
$B\sim 10^{10}-10^{11}$~G; these NSs are the so-called central compact objects
(CCOs), i.e., NSs near the centre of their respective supernova remnants
(SNRs; see, e.g., \citealt{deluca08,gotthelfhalpern08}, for review;
see also \citealt{halperngotthelf10}).
In particular, X-ray timing measurements reveal pulsations and spin-down
rates that imply from eq.~(\ref{eq:pdot}) that
$\Bp<4.0\times 10^{11}$~G for PSR~J0821$-$4300 in SNR Puppis~A
\citep{gotthelfetal10},
$\Bp<6.6\times 10^{11}$~G for 1E~1207.4$-$5209 in SNR PKS~1209$-$51/52
(also known as G296.5+10.0; \citealt{gotthelfhalpern07}), and
$\Bp=6.1\times 10^{10}$~G for PSR~J1852$+$0040 in SNR Kes~79
\citep{halperngotthelf10};
hereafter, we refer to these three CCOs as Puppis~A, 1E~1207, and Kes~79,
respectively.
In addition to timing measurements, the X-ray spectrum of 1E~1207 shows
absorption features which could be due to electron cyclotron resonance at
$B\approx 7-8\times 10^{10}$~G \citep{sanwaletal02,mereghettietal02b},
while a possible line in the spectrum of PSR~J0821$-$4300 could indicate
$B\approx 8-10\times 10^{10}$~G \citep{gotthelfhalpern09,suleimanovetal10}.
Finally, although pulsations have not been detected in another member of this
class, the NS in SNR Cassiopeia~A
\citep{murrayetal02,mereghettietal02,ransom02,pavlovluna09,halperngotthelf10},
X-ray spectral fits suggest $B<10^{11}$~G
\citep{hoheinke09,heinkeho10,shterninetal11}.

In this work, we show that young ($\le 10^4$~yr) NSs, like the CCOs,
can be understood as NSs which are born with $B\gtrsim 10^{11}$~G but whose
fields have been buried deep beneath the surface \citep{romani90},
perhaps by a post-supernova episode of hypercritical accretion
\citep{chevalier89,geppertetal99,bernaletal10}.
These fields then diffuse to the surface on the timescale of $10^3-10^4$~yr,
so that only now do we see a surface $B\sim 10^{10}-10^{11}$~G.
It is important to emphasize that we are primarily concerned with
field {\it growth} in $< 10^4$~yr, which is much shorter than the
estimate given by eq.~(\ref{eq:tohm}).
We briefly summarize relevant past works.
\citet{youngchanmugam95} studied field growth in old NSs (with low
temperatures) with buried magnetic fields and found that the
surface field is restored after $\sim 10^{10}$~yr if the total accreted mass
is $\lesssim 0.04M_\odot$.
\citet{muslimovpage95} examined field evolution when the field (confined
to the crust) is buried at shallow depths, where the density is
$\denssub\sim 5\times 10^{10}$ and $4.3\times 10^{11}\mbox{ g cm$^{-3}$}$;
as a result of the shallow submergence, the diffusion times are short
($\mbox{a few}\times 10^2-10^3$~yr).
\citet{muslimovpage96} modelled field evolution in three young ($<1900$~yr),
fast rotating ($P = 30$, 50, and 150~ms and initial spin period
$P_\mathrm{i} = 20-40$~ms) radio pulsars with a measured braking index
[see eq.~(\ref{eq:brakeindex})],
taking into account cooling of the NS, which influences the
electrical conductivity and hence the field diffusion timescale
[see eq.~(\ref{eq:tohm})].
The magnetic flux is assumed to be frozen into the core on timescales
$\lesssim 10^6$~yr.
Only shallow submergence
[$\denssub\approx (1-3)\times 10^{10}\mbox{ g cm$^{-3}$}$]
is considered since the pulsars require rapid field growth
(to $B\sim 10^{13}$~G) given their young age;
note that the rapid field growth also results in a more rapid spin-down rate.
Finally,
\citet{geppertetal99} considered field growth of a purely crustal field
at three submergence depths,
$\denssub=10^{12}, 10^{13}, 10^{14}\mbox{ g cm$^{-3}$}$.

Discoveries (especially the recognition of the CCO class of NSs)
since the works discussed above, as well as advancements in the
theory of the relevant electrical and thermal conductivities,
motivate the current work.
We follow a similar methodology as in the previous works that examined
magnetic field growth (see \citealt{geppertetal99}, and references therein),
but here we use updated physics (see Sec.~\ref{sec:econd} and \ref{sec:evolt})
and parameters that are particular to the properties of the CCOs.
By comparing the measured properties of specific CCOs with our calculations,
we obtain constraints on the birth magnetic field and submergence depth
in each NS.
The commonality of these two values in the various sources further
strengthens the distinctiveness of this class of NSs.

In Section~\ref{sec:model}, we discuss the magnetic field evolution
equation and the input physics, in particular, the conductivities,
equation of state, and temperature evolution.
In Section~\ref{sec:results}, we describe our general results.
In Section~\ref{sec:cco}, we apply our results to the three CCOs with
known spin periods and period derivatives or magnetic fields, in order to
determine the strength of their birth magnetic fields and depth
of submergence.
We summarize our results and discuss their implications in
Section~\ref{sec:discuss}.

\section{Neutron star model} \label{sec:model}

\subsection{Magnetic field evolution} \label{sec:evolb}

Our calculation of magnetic field evolution is based on
\citet{urpinmuslimov92}.
To determine the evolution of the submerged field,
we solve the induction equation
\be
\frac{\partial\vecB}{\partial t} =
 -\nabla\times\left(\frac{c^2}{4\pi\econd}\nabla\times\vecB\right)
 = -\nabla\times\left(\bdiff\nabla\times\vecB\right), \label{eq:induct1}
\ee
where $\bdiff\equiv c^2/4\pi\econd$ is the magnetic diffusivity.
Our primary interest is in the NS crust, which is predominantly in a solid
state (see below); therefore we neglect internal fluid motion.
We assume the magnetic field in the stellar interior is dipolar
and given by the simple azimuthal vector potential,
$\vecA=\Bstar R^2s(r,t)\sin\theta/r\,\hat{\phi}$, so that the axially
symmetric field is given by
\begin{eqnarray}
\vecB &=& \Bstar R^2\left(\frac{2s}{r^2}\cos\theta\hat{r}
 -\frac{1}{r}\frac{\partial s}{\partial r}\sin\theta\hat{\theta}\right)
 \nonumber\\
&=& \Bstar\left(b_r\cos\theta\hat{r}
 + b_\theta\sin\theta\hat{\theta}\right), \label{eq:magb}
\end{eqnarray}
where $\Bstar $ is the ``birth'' magnetic field
and the second equality defines the normalized fields $b_r$ and $b_\theta$.
The birth magnetic field is understood to be the surface magnetic field
after NS formation but prior to mass accretion; accretion then buries
and compresses the birth field.
Assuming only radial variations of the electrical conductivity,
i.e., $\econd=\econd(r)$,
eq.~(\ref{eq:induct1}) reduces to the one-dimensional equation
\be
\frac{\partial s}{\partial t}
 = \bdiff\left(\frac{\partial^2s}{\partial r^2}-\frac{2s}{r^2}\right),
 \label{eq:induct2}
\ee
and the field evolution is solely determined by the electrical conductivity.

We assume that the accreted material is non-magnetic, so that the
initial field decreases very rapidly at shallower depths than $\denssub$.
A pre-existing field at these depths would shorten the growth time;
however this field must be ordered, whereas (hypercritical) accretion is
likely to be turbulent \citep{chevalier89,geppertetal99,bernaletal10}.
The boundary conditions are then
\be
\frac{\partial s}{\partial r}+\frac{s}{R} = 0
\ee
at the surface $r=R$ and $s\rightarrow\mbox{constant}$ in the deep interior.
The constant determines the two internal configurations for the initial
magnetic field that we consider (see thick lines in Fig.~\ref{fig:profile}):
(1) confined crustal field, i.e.,
the diffusion of the magnetic field does not extend into the core
(e.g., as a result of a superconducting core) and
the field is negligible throughout the star except at the submergence
density $\denssub$ or position $\rsub$, where
\be
s(r,0) = (1/2)e^{-[(r-\rsub)/\rsubwidth]^2} \label{eq:bc1}
\ee
and $\rsubwidth=10^{-3}\rsub$;
(2) deep crustal ``core'' field, i.e.,
the field is finite at $\rho>\denssub$ or
\be
s(r,0) = \frac{1}{2}\left\{\begin{array}{ll}
e^{-[(r-\rsub)/\rsubwidth]^2} & \mbox{for } \rho<\denssub \\
1 & \mbox{for } \rho>\denssub
\end{array}\right. . \label{eq:bc2}
\ee
The Ohmic diffusion timescale [see eq.~(\ref{eq:tohm})] near the core
is much longer than the times under consideration here ($t\le 10^4$~yr),
so that a constant core field is valid;
also field decay and flux expulsion due to superfluid motion
in the core only occurs at $t>10^4$~yr since the low surface magnetic field
produces little spin-down \citep{konenkovgeppert01}.
Note that the factor of 1/2 in eqs.~(\ref{eq:bc1}) and (\ref{eq:bc2})
has been included so that the normalized magnetic field is
$b_r(\denssub)\approx 1$ and $B(\denssub)\approx\Bstar$
[see eq.~(\ref{eq:magb})].

\subsection{Electrical conductivity} \label{sec:econd}

The dominant contribution to the electrical conductivity
depends on whether the matter is in a liquid or solid state.
The melting temperature $T_\mathrm{m}$ is given by
\be
T_\mathrm{m}=3.04\times 10^7\mbox{ K }(Z/26)^{5/3}(170/\Gamma)x,
 \label{eq:tmelt}
\ee
where $x=(Z\rho_6/A)^{1/3}$,
$Z$ and $A$ are the charge number and mass number of the ion, respectively,
$\Gamma=Z^2e^2/a\kb T$, $a=(3/4\pi n_\mathrm{i})^{1/3}$,
$n_\mathrm{i}$ is the ion number density, and
$\rho_6=\rho/10^6\mbox{ g cm$^{-3}$}$.
The electrical conductivity is mainly determined by electron-ion (ei)
scattering when $T>T_\mathrm{m}$ and electron-phonon (ph) scattering
when $T<T_\mathrm{m}$.
For illustrative purposes, these are given by
\begin{eqnarray}
\econdei &=& 8.53\times 10^{21}\mbox{ s$^{-1}$ }
 \frac{x^3}{Z\Lambda_\mathrm{ei}(1+x^2)} \label{eq:econdei} \\
\econdph &=& 1.21\times 10^{28}\mbox{ s$^{-1}$ }
 \frac{\sqrt{u^2+0.017}}{Tu}\frac{x^4}{2+x^2}, \label{eq:econdph}
\end{eqnarray}
where $\Lambda_\mathrm{ei}$ is the Coulomb logarithm,
$u=0.45T/T_\mathrm{D}$, and $T_\mathrm{D}$ is the Debye temperature
\be
T_\mathrm{D} = 3.4\times 10^6\mbox{ K }(Z/A)^{1/2}x^{3/2}. \label{eq:tdebye}
\ee
Equations~(\ref{eq:econdei}) and (\ref{eq:econdph})
(\citealt{yakovlevurpin80}; see also \citealt{itohetal93}, and references
therein) were used by all previous works to solve the induction
equation in order to study magnetic field growth (see Sec.~\ref{sec:intro}).
Instead, we use here
{\small CONDUCT08}\footnote{http://www.ioffe.ru/astro/conduct/},
which implements the latest advancements in calculating the conductivities,
including electron-electron contributions
\citep{potekhinetal99,cassisietal07,chugunovhaensel07};
we also tested {\small SFITTING} (N.~Itoh, private comm.; \citealt{itohetal08}).
We assume no contribution by impurity scattering, which only becomes
important at high densities and low temperatures,
and no magnetic field effects on the conductivities.
At high magnetic fields and low temperatures, electron motion is strongly
influenced by the magnetic field (see, e.g., \citealt{yakovlevkaminker94}).
As a result, the electrical and thermal conductivities become anisotropic,
depending on whether the motion is parallel or transverse to the direction
of the field \citep{potekhin99}.
We only consider here young NSs with high interior temperatures;
thus magnetic field effects are minimal since the field is in the
non- or weakly-quantizing regime, i.e., $T>T_\mathrm{B}$, where
\be
T_\mathrm{B} = 1.34\times 10^8\mbox{ K }(B/10^{12}\mbox{ G})(1+x^2)^{-1/2},
 \label{eq:tmag}
\ee
and $\rho\gg\rho_\mathrm{B}
=2.228\times 10^{5}\mbox{ g cm$^{-3}$}(A/Z)(B/10^{13}\mbox{ G})^{3/2}$.
However, magnetic field effects can be important in the cooler, low-density
atmosphere, which produces the observed X-ray emission from NSs
(see, e.g., \citealt{geppertetal04,ponsetal09}).

\subsection{Equation of state} \label{sec:eos}

\begin{figure}
\resizebox{\hsize}{!}{\includegraphics{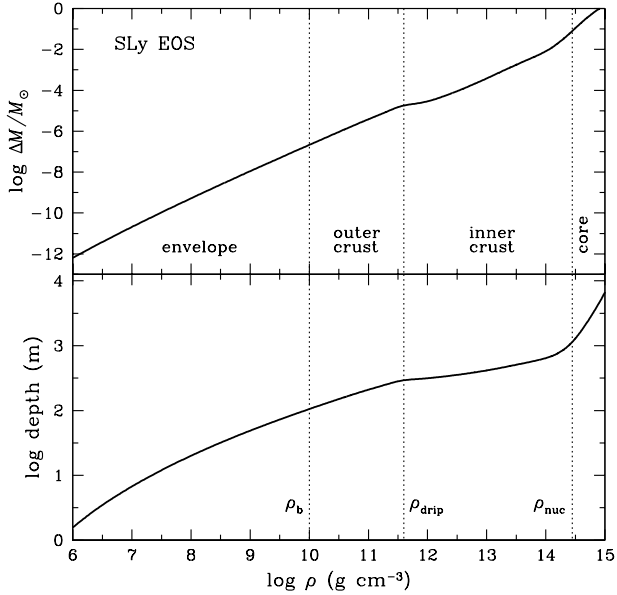}}
\caption{
Relative (accreted) mass $\Delta M=M-m(r)$ and depth $R-r$ as a function
of density for a neutron star with the SLy EOS.
Dotted lines indicate the boundary between the envelope and crust
(at $\rho_\mathrm{b}\approx 10^{10}\mbox{ g cm$^{-3}$}$),
the boundary between the outer and inner crust
(at $\rho_\mathrm{drip}\approx 4\times 10^{11}\mbox{ g cm$^{-3}$}$),
and the boundary between the crust and core
(at $\rho_\mathrm{nuc}\approx 2.8\times 10^{14}\mbox{ g cm$^{-3}$}$).
}
\label{fig:acc}
\end{figure}

The electrical conductivity depends on the local density and temperature
(as well as the chemical composition, which we take to be iron throughout
for simplicity);
therefore the NS density and temperature (radial) profiles must be known.
We calculate the density profile by solving the TOV stellar structure
equations \citep[see, e.g.,][]{shapiroteukolsky83}, supplemented by an
equation of state (EOS) describing the pressure $P$ as a function of density
(see, e.g., \citealt{haenseletal07,lattimerprakash07}, for review).
We use the analytic fits by \citet{haenselpotekhin04} of the
SLy EOS, which is a moderately-stiff EOS with a maximum mass of
$2.05\,M_\odot$ for the central density
$\rho_\mathrm{c}=2.9\times 10^{15}\mbox{ g cm$^{-3}$}$
\citep{douchinhaensel01}.
The NS we consider has $\rho_\mathrm{c}=1.2\times 10^{15}\mbox{ g cm$^{-3}$}$,
mass $M=1.63\,M_\odot$, and radius $R=11.5\mbox{ km}$;
the inner crust
(at $\rho<\rho_\mathrm{nuc}\approx 2.8\times 10^{14}\mbox{ g cm$^{-3}$}$)
is at depths less than $\approx 1.1\mbox{ km}$;
the outer crust
(at $\rho<\rho_\mathrm{drip}\approx 4\times 10^{11}\mbox{ g cm$^{-3}$}$)
is at depths less than $\approx 290\mbox{ m}$;
the heat-blanketing envelope
(at $\rho<\rho_\mathrm{b}\approx 10^{10}\mbox{ g cm$^{-3}$}$; see below)
is at depths less than $\approx 100\mbox{ m}$.
The boundaries of the inner and outer crust and envelope are shown in
Fig.~\ref{fig:acc}, as well as the mass above a certain depth
or density $\Delta M$ [$\equiv M-m(r)$,
where $m(r)$ is the stellar mass enclosed within radius $r$].
$\Delta M$ is an indication of the amount of accreted mass needed
to submerge the magnetic field down to a given density
(see, e.g., \citealt{geppertetal99},
for EOS-dependence of $\Delta M$ and depth).

We only consider a single theoretical nuclear EOS and the standard
mechanism (modified Urca process) for neutron star cooling (see below).
Varying the EOS would result in several effects on the Ohmic diffusion/decay
timescale $\tohm\propto\econd(T) L^2$ [see eq.~(\ref{eq:tohm})],
in particular, a change to the thickness of the crust, as well as
possibly inducing more rapid neutrino cooling for high mass NSs.
EOS effects have been studied in the context of long-term field decay
(see, e.g., \citealt{urpinkonenkov97,konenkovgeppert01}),
while fast neutrino cooling was needed to explain two of the three
young pulsars in \citet{muslimovpage96}.
For the CCOs, fast neutrino cooling is not required, and the effects of
different EOSs on the field evolution are beyond the scope of this work.

\subsection{Temperature evolution} \label{sec:evolt}

Neutrino emission is the main source of cooling during the first
$\sim 10^5\mbox{ yr}$ after NS formation.
The thermal history of a NS is primarily determined by the neutrino
luminosity and heat capacity of the core and the composition
(i.e., thermal conductivity) of the surface layers
(see \citealt{tsuruta98,yakovlevpethick04,pageetal06}, for review)
At very early times, the core cools via neutrino emission while
the temperature of the thermally-decoupled crust remains nearly constant.
A cooling wave travels from the core to the surface, bringing the NS to a
relaxed, isothermal state.  Depending on the properties of the crust,
the relaxation time can take $\sim 10-100\mbox{ yr}$
\citep{lattimeretal94,gnedinetal01}.
For the next $\sim 10^5-10^6\mbox{ yr}$, surface temperature changes reflect
changes in the interior temperature as neutrino emission continuously
removes heat from the star.

For our temperature evolution, we take the NS to cool by the standard
(slow-cooling) modified Urca process of neutrino emission, which results
in $T\propto t^{-1/6}$ \citep{tsuruta98,yakovlevpethick04,pageetal06}.
Specifically, we use \citep{yakovlevetal11}
\begin{eqnarray}
T(r,t) &=& 9.07\times 10^8\mbox{ K }
 e^{-\phi(r)}\left(1-\frac{2GM}{c^2R}\right) \nonumber\\
 && \times \left(1+0.12R_6^2\right)\left(\frac{\mbox{1 yr}}{t}\right)^{1/6},
 \label{eq:evolt}
\end{eqnarray}
where $\phi(r)$ is the metric function that determines the gravitational
redshift.
For our assumed NS model (see Sec.~\ref{sec:eos}), the initial core
temperature
$\Tc(t=0)\equiv T(0,0)=1.1\times 10^9\mbox{ K}$.
Faster cooling processes, such as the direct Urca process, would cause
the temperature to decrease more rapidly. This can lead to an increase
in the electrical conductivity [see eq.~(\ref{eq:econdph})], which
could slow magnetic field evolution since the Ohmic diffusion timescale
increases.
However our results indicate that cooling beyond modified Urca is
not needed.

\begin{figure}
\resizebox{\hsize}{!}{\includegraphics{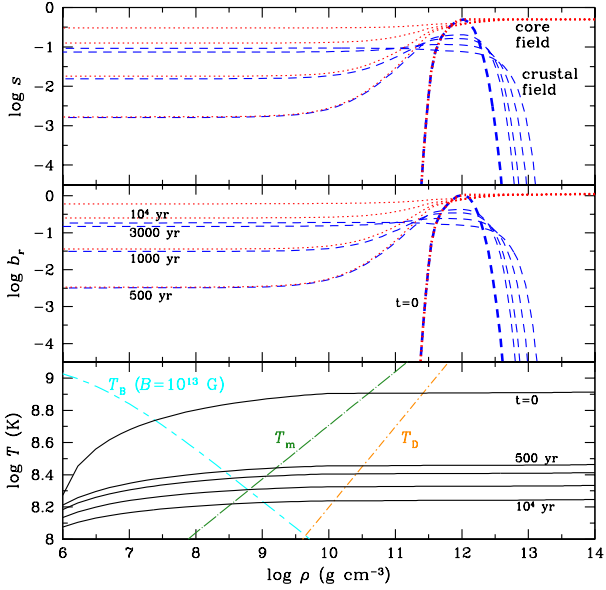}}
\caption{
Profiles of $s$, radial component of normalized magnetic field $b_r$, and
temperature $T$ at five time steps ($t=0$, 500, 1000, 3000, and $10^4$~yr)
for a magnetic field that is initially submerged at a density
$\denssub=10^{12}\mbox{ g cm$^{-3}$}$
Dashed and dotted lines are for the crustal and core field configurations,
respectively (see text);
the temperature profile (solid lines) is the same for both configurations.
$T_\mathrm{B}$ (short-long-dashed line), $T_\mathrm{m}$
(long-dashed-dotted line) and $T_\mathrm{D}$ (short-dashed-dotted line)
are the magnetic (at $B=10^{13}$~G), melting, and Debye temperatures,
respectively.
}
\label{fig:profile}
\end{figure}

\begin{figure}
\resizebox{\hsize}{!}{\includegraphics{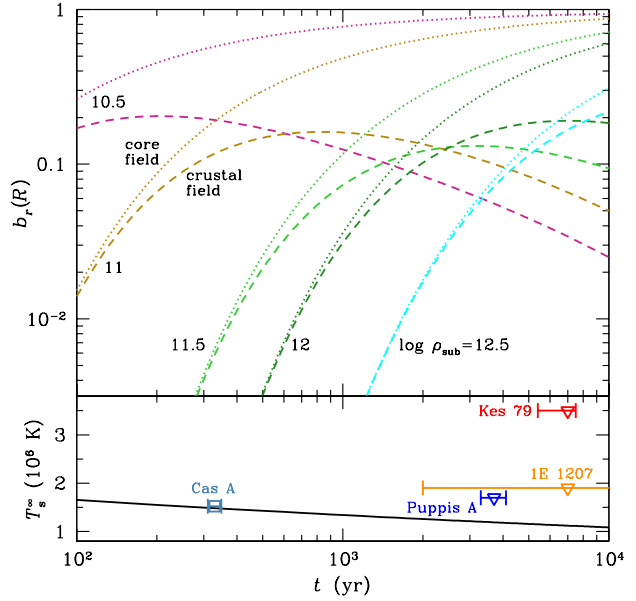}}
\caption{
Evolution of the normalized magnetic field at the surface $b_r(R)$ and
redshifted surface temperature $\Ts^\infty$ for submergence densities
$\log\denssub(\mbox{g cm$^{-3}$})=10.5,11,11.5,12,12.5$.
Dashed and dotted lines are for the crustal and core field configurations,
respectively (see text);
$\Ts^\infty$ (solid line) is the same for both configurations.
Triangles are the upper limits on the temperature of the entire neutron star
surface of the CCOs Puppis~A, 1E~1207, and Kes~79,
and the square is the surface temperature of the CCO in Cas~A
(see text and Table~\ref{tab:cco}).
}
\label{fig:evolb}
\end{figure}

Since we are concerned with NSs that are $> 100\mbox{ yr}$ old, we assume
that the NS has an isothermal core, specifically
$T(r)e^{\phi(r)}=\mbox{constant}$.  On the other hand, the outer layers
(i.e., envelope) serve as a heat blanket, and there exists a temperature
gradient from core temperatures $\Tc\sim 10^8-10^9\mbox{ K}$ to surface
temperatures $\Ts\sim 10^5-10^6\mbox{ K}$.
We combine the equations of hydrostatic equilibrium,
\be
\frac{dP}{dr} = -\rho\gs,
\ee
where $\gs=(1-2GM/c^2R)^{-1/2}GM/R^2$ is the surface gravity
($\gs=2\times 10^{14}\mbox{ cm s$^{-2}$}$ for our assumed NS model;
see Sec.~\ref{sec:eos}),
and thermal diffusion,
\be
\frac{dT}{dr} = -\frac{F}{\tdiff} = -\frac{\sigsb\Ts^4}{\tdiff},
\ee
where $\tdiff$ is the thermal conductivity,
to obtain \citep{gudmundssonetal82}
\be
\frac{dT}{d\rho} = \frac{\sigsb}{\rho\tdiff}\frac{\Ts^4}{\gs}\frac{dP}{d\rho}
 = \frac{3}{16}\frac{\opact}{\gs}\frac{\Ts^4}{T^3}\frac{dP}{d\rho},
\label{eq:tempode}
\ee
where $\opact = 16\sigsb T^3/3\rho\tdiff$ is the opacity due to conduction
and radiation, i.e., $1/\opact=1/\opacr+1/\opacc$.
We consider the radiative opacity to be due to free-free absorption
$\opacff\approx 6\times 10^{22}\mbox{ cm$^2$ g$^{-1}$ }Z^3A^{-2}\rho T^{-7/2}$
and electron scattering $\opaces = Z\sigth/Am_\mathrm{p}$.
The thermal conductivity is calculated using {\small CONDUCT08}
(see Sec.~\ref{sec:econd}).
Note that we ignore relativistic effects in the derivation of
eq.~(\ref{eq:tempode}) since uncertainties in the input physics exceed
the effects of their inclusion on the results \citep{gudmundssonetal82}.
For simplicity,
we assume the magnetic field does not determine the surface temperature
distribution (see Sec~\ref{sec:econd}).
\citet{changbildsten04,changetal10} showed that nuclear burning would very
rapidly remove any surface light elements for the high temperatures present
in young NSs; therefore we only consider a surface composed of iron.
The relation between the interior and surface temperatures is then given by
\citep{gudmundssonetal82}
\be
\Ts = 8.701\times 10^5\mbox{ K }g_{14}^{1/4}
 \left\{\frac{T[r(\rho_\mathrm{b}),t]}{10^8\mbox{K}}\right\}^{11/20}.
\ee
Equation~(\ref{eq:tempode}) is solved from $\rho=\rho_\mathrm{b}$ to
$10^6\mbox{ g cm$^{-3}$}$ for a given $T[r(\rho_\mathrm{b}),t]$
from eq.~(\ref{eq:evolt});
we take $\rho_\mathrm{b}=10^{10}\mbox{ g cm$^{-3}$}$
\citep[see, e..g,][]{gudmundssonetal82,yakovlevpethick04}.
The evolution of the temperature profile is shown in Fig.~\ref{fig:profile},
while the evolution of the redshifted surface temperature
$\Ts^\infty$ [$=\Ts(1-2GM/c^2R)^{1/2}$] is shown in Fig.~\ref{fig:evolb}.
Note that the temperature profile is independent of the magnetic field
or its submergence depth or evolution.

\section{Results} \label{sec:results}

Here we give examples of our solution of the induction equation, which
illustrates the evolution of a buried crustal or core magnetic field.
Figure~\ref{fig:profile} shows the interior profiles of $s$ and the
normalized magnetic field $b_r$ [$=\Bp(r)/\Bstar$] for the case where
the field is initially submerged at a density
$\denssub=10^{12}\mbox{ g cm$^{-3}$}$.
Note that $\denssub=10^{12}\mbox{ g cm$^{-3}$}$ implies an accreted
mass $\Delta M\approx 3\times 10^{-5}M_\odot$ (see Fig.~\ref{fig:acc}).
For the crustal field configuration, one can see that the peak in the
magnetic field at $\denssub$ at $t=0$ has decreased due to diffusion/decay
by about an order of magnitude in $10^4$~yr,
so that $s$ and $b_r$ are roughly constant as a function of density at
$\rho<\denssub$;
the field also diffuses to greater depths ($\rho>\denssub$) as the
evolution continues.

Figure~\ref{fig:evolb} shows the evolution of the normalized field at
the surface $b_r(R)$ for crustal and core fields and various submergence
densities.
As found in previous works which considered crustal fields
\citep{muslimovpage95,geppertetal99}, the shallower the initial field is
buried, the more quickly the surface field grows to a maximum.
Because the peak surface field occurs at later times as the submergence
depth increases, Ohmic field decay causes the value of the surface field
$b_r(R)$ to decrease with $\denssub$
(for $\denssub\lesssim 10^{12}\mbox{ g cm$^{-3}$}$ and $t\lesssim 10^4$~yr);
thus a stronger birth magnetic field $\Bstar$ is needed to yield
the same (observed) $B$.
In the core field configuration, the surface field increases
(and saturates) earlier with shallower burial.
It is evident from Fig.~\ref{fig:profile} and \ref{fig:evolb} that the
early growths of the magnetic field are very similar in the crustal and
core field configurations.
At later times, the surface field decays, and the maximum field is lower in
the crustal configuration than in the core configuration.
Thus to match an observed $B$, a lower birth field $\Bstar$ is
required in the core configuration.
Finally, we note that the growth times shown here are shorter than those
seen in previous works \citep[see, e.g.,][]{muslimovpage95,geppertetal99}
due to our use of the improved conductivities (see Sec.~\ref{sec:econd}),
and the difference increases the deeper the field is buried.

\section{Comparison to CCOs} \label{sec:cco}

\subsection{Properties of observed CCOs}

We now consider the three CCOs with measured spin periods $P$ and either
spin period derivative $\Pdot$ or magnetic field (at the pole)
$\Bp^\mathrm{line}$ (see Table~\ref{tab:cco}).
Note that $\Pdot$ is obtained from X-ray timing, while
$\Bp^\mathrm{line}$ is determined from X-ray spectra,
and both of these measurements contain systematic uncertainties.
In particular, eq.~(\ref{eq:pdot}) assumes a dipolar magnetic field
radiating into a vacuum.
\citet{spitkovsky06} calculated a magnetospheric-equivalent equation
that yields $\Bp$ lower by a factor of $<1.7$.
On the other hand, the relation between spectral features and the NS
magnetic field is not definitive.

\begin{table}
\caption{Central Compact Objects \label{tab:cco}}
\begin{tabular}{c c c c}
\hline
 & Puppis A & 1E~1207 & Kes 79 \\
\hline
Age (kyr) & 3.7$\pm$0.4 & $\sim 7^a$ & 5.4$-$7.5 \\
$P$ (ms) & 112 & 424 & 105 \\
$\Pdot$ ($10^{-17}$ s s$^{-1}$) & $<35$ & $<25$ & $0.868\pm 0.009$ \\
$\Bp$ ($10^{11}$ G) & $<4.0$ & $<6.6$ & 0.61 \\
$\Bp^\mathrm{line}$ ($10^{11}$ G) & 0.8$-$0.9 & 0.7$-$0.8 & --- \\
$\Edot$ ($10^{33}$ ergs s$^{-1}$) & $<10$ & $<0.13$ & 0.30 \\
$\tau_\mathrm{c}$ ($10^6$ yr) & $>5.1$ & $>27$ & 192 \\
$\Ts^\infty$ ($10^6$ K) & $<1.7$ & $<1.9$ & $<3.5$ \\
References & 1,2,3 & 4,5,6 & 7,8 \\
\hline
\end{tabular}
\begin{quote}
{\scshape Notes:}\\
$^a$Age estimate is uncertain by a factor of three \citep{rogeretal88}.
References:
(1) \citet{winkleretal88},
(2) \citet{gotthelfhalpern09},
(3) \citet{gotthelfetal10},
(4) \citet{sanwaletal02},
(5) \citet{delucaetal04},
(6) \citet{gotthelfhalpern07},
(7) \citet{sunetal04},
(8) \citet{halperngotthelf10}.
\end{quote}
\end{table}

For Puppis~A,
\citet{gotthelfhalpern09} find the best-fit model for the X-ray spectrum
includes the addition of a Gaussian emission line at $0.79\pm 0.02$~keV,
though a good fit can also be obtained without the line.
Alternatively, \citet{suleimanovetal10} suggest the spectral feature can
be interpreted as an absorption line at 0.9~keV.
For our purposes, we only require knowledge of the line energy,
and we use the 0.8~keV value.
For 1E~1207, we interpret the absorption lines seen in 1E~1207 at 0.7 and
1.4~keV \citep{sanwaletal02,mereghettietal02b,bignamietal03} as being due
to the electron cyclotron resonance \citep{potekhin10,suleimanovetal10}.
Spectral features due to the electron cyclotron resonance occur at
\be
E_\mathrm{c,e} = \hbar eB/m_\mathrm{e}c
 = 1.158\mbox{ keV }(B/10^{11}\mbox{ G}),
\ee
and we assume a gravitational redshift factor $(1-2GM/c^2R)^{-1/2}$
in the range $1.2-1.35$.
For the two CCOs with spectrally-measured magnetic fields and limits on
$\Pdot$, the $\Bp^\mathrm{line}$ satisfies the upper bound set by $\Pdot$.

Table~\ref{tab:cco} also shows
several related parameters that are conventionally given for a pulsar with
measured $P$ and $\Pdot$: rotational energy loss rate
\be
\Edot = 4\pi^2 I\frac{\Pdot}{P^3} = 2\pi^2\gamma I\frac{\Bp^2}{P^4},
 \label{eq:edot}
\ee
characteristic age
\be
\tau_\mathrm{c} = \frac{P}{2\Pdot} = \frac{P^2}{\gamma\Bp^2}, \label{eq:tc}
\ee
and braking index $n$
\be
n = 3-2\frac{\Bdot}{\Bp}\frac{P}{\Pdot}
 = 3-\frac{4}{\gamma}\frac{\Bdot}{\Bp^3}P^2. \label{eq:brakeindex}
\ee
Note that there is a selection effect of detecting sources with short
periods since $\Edot\propto P^{-3}$.
The characteristic age is much longer than the true age if the
pulsar is born with a initial period $P_\mathrm{i}$ near its currently
observed period $P$.  Since the observed $\Pdot$ is low for the CCOs,
there is little spin period evolution, and $\tau_\mathrm{c}\gg 10^4$~yr.
Also given in Table~\ref{tab:cco} and plotted in Fig.~\ref{fig:evolb} are
the blackbody temperatures obtained from {\it XMM-Newton} observations
of Puppis~A, 1E~1207, and Kes~79;
the temperature for Puppis~A is an upper limit \citep{gotthelfetal10},
while the temperatures for 1E~1207 and Kes~79 are the cooler component
of the two-temperature spectral fits \citep{delucaetal04,halperngotthelf10}
and represent upper limits on the temperature of the entire NS surface.
Figure~\ref{fig:evolb} also plots
the temperature obtained from spectral fits to {\it Chandra}
observations of the CCO in the Cassiopeia~A (Cas~A) supernova remnant
using model atmosphere spectra \citep{hoheinke09,yakovlevetal11};
note that this temperature is for the entire NS surface,
and the temperature has been seen to decrease by $\approx 4\%$ over
the last 11~yr due to the cooling of the NS
\citep{heinkeho10,shterninetal11}.

\subsection{CCOs as individual sources} \label{sec:ccothree}

We can now use our evolution calculations to constrain the birth
magnetic field $\Bstar$ and submergence density $\denssub$.
We begin by integrating eq.~(\ref{eq:pdot}) to obtain
\be
P(t) = \left[P_\mathrm{i}^2
 + \gamma\int_{t_{\mathrm{i}}}^t\Bp(t')^2\,dt'\right]^{1/2},
\ee
which allows us to determine $\Pdot(t)$.
We then calculate $\Bdot$, $\Edot$, $\tau_\mathrm{c}$, and $n$.

\begin{figure}
\resizebox{\hsize}{!}{\includegraphics{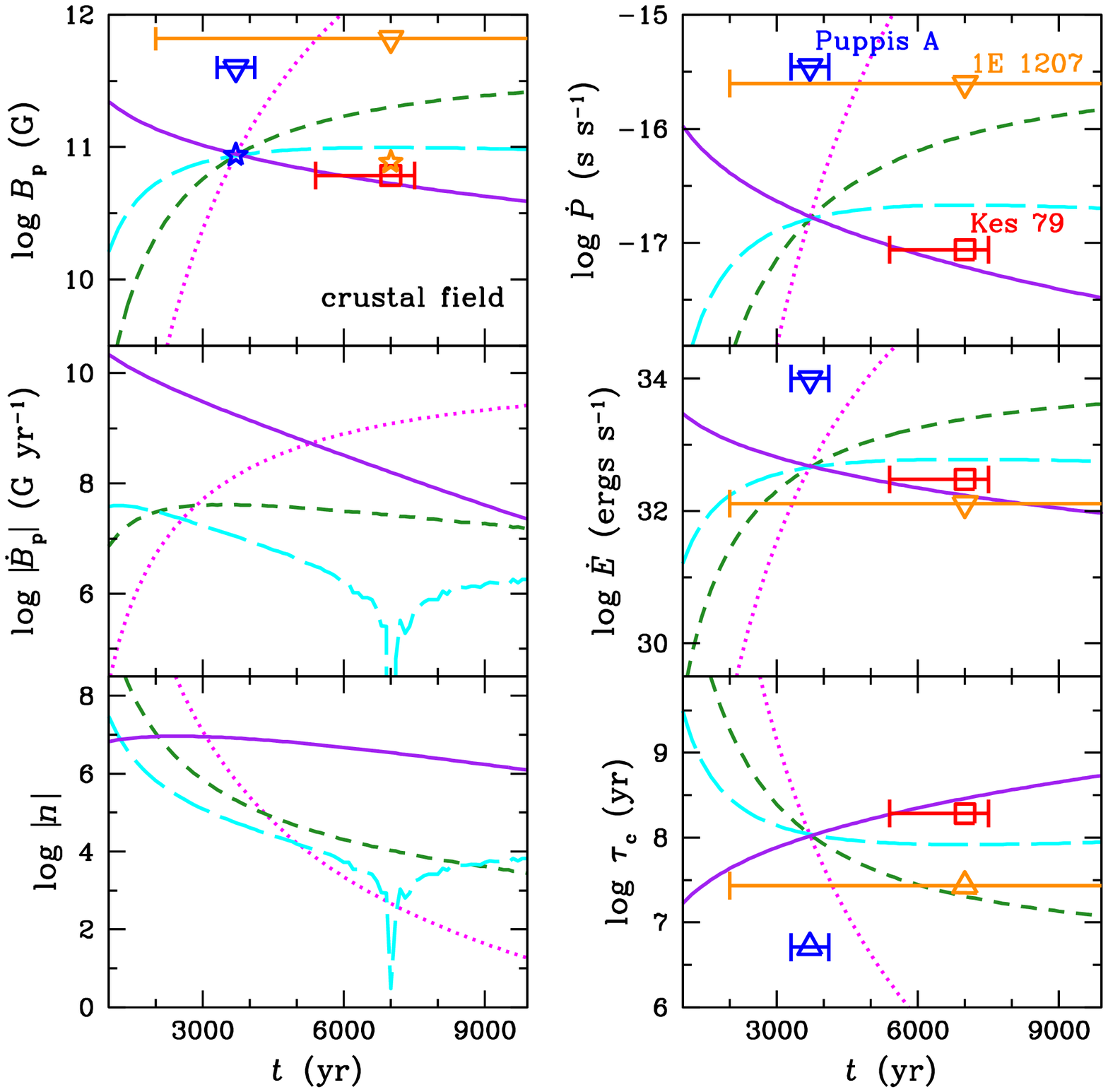}}
\caption{
Pulsar observables: magnetic field at the pole $\Bp$,
spin period derivative $\Pdot$,
rotational energy loss rate $\Edot$.
and characteristic age $\tau_\mathrm{c}$
for the CCOs Puppis~A, 1E~1207, and Kes~79;
squares are values obtained from timing measurements,
triangles are upper/lower limits obtained from timing, and
stars are from spectral measurements (see Table~\ref{tab:cco}).
Lines are evolution models with an initial spin period
$P_\mathrm{i}=0.112$~s and buried crustal field at density $\denssub$
and birth magnetic field $\Bstar$:
$[\log\denssub\mbox{(g cm$^{-3}$)},\Bstar\mbox{(G)}]=$
(8.0,$2.8\times 10^{13}$; solid),
(12.0,$5.2\times 10^{11}$; long-dashed),
(12.5,$1.2\times 10^{12}$; short-dashed), and
(13.0,$1.5\times 10^{15}$; dotted).
Also shown are the calculated magnetic field derivative $\Bdot$ and
braking index $n$; the singularity is because of a change in sign of
$\Bdot$, from a growing to decaying $\Bp$.
}
\label{fig:obs_pupa}
\end{figure}

\begin{figure}
\resizebox{\hsize}{!}{\includegraphics{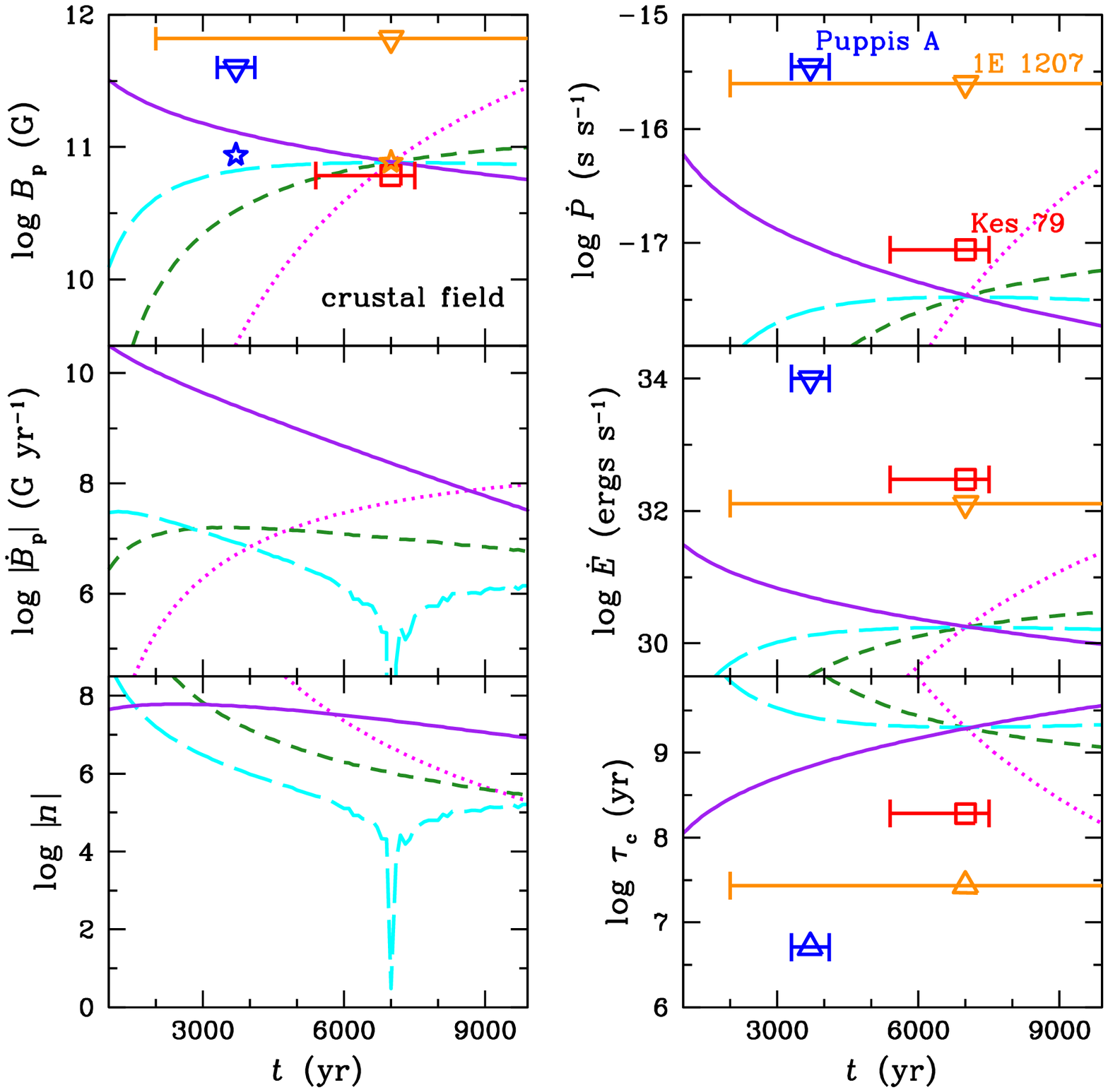}}
\caption{
Pulsar observables: magnetic field at the pole $\Bp$,
spin period derivative $\Pdot$,
rotational energy loss rate $\Edot$.
and characteristic age $\tau_\mathrm{c}$
for the CCOs Puppis~A, 1E~1207, and Kes~79;
squares are values obtained from timing measurements,
triangles are upper/lower limits obtained from timing, and
stars are from spectral measurements (see Table~\ref{tab:cco}).
Lines are evolution models with an initial spin period
$P_\mathrm{i}=0.424$~s and buried crustal field at density $\denssub$
and birth magnetic field $\Bstar$:
$[\log\denssub\mbox{(g cm$^{-3}$)},\Bstar\mbox{(G)}]=$
(8.0,$4.1\times 10^{13}$; solid),
(12.0,$4.0\times 10^{11}$; long-dashed),
(12.5,$4.6\times 10^{11}$; short-dashed), and
(13.0,$5.6\times 10^{13}$; dotted).
Also shown are the calculated magnetic field derivative $\Bdot$ and
braking index $n$; the singularity is because of a change in sign of
$\Bdot$, from a growing to decaying $\Bp$.
}
\label{fig:obs_1e1207}
\end{figure}

\begin{figure}
\resizebox{\hsize}{!}{\includegraphics{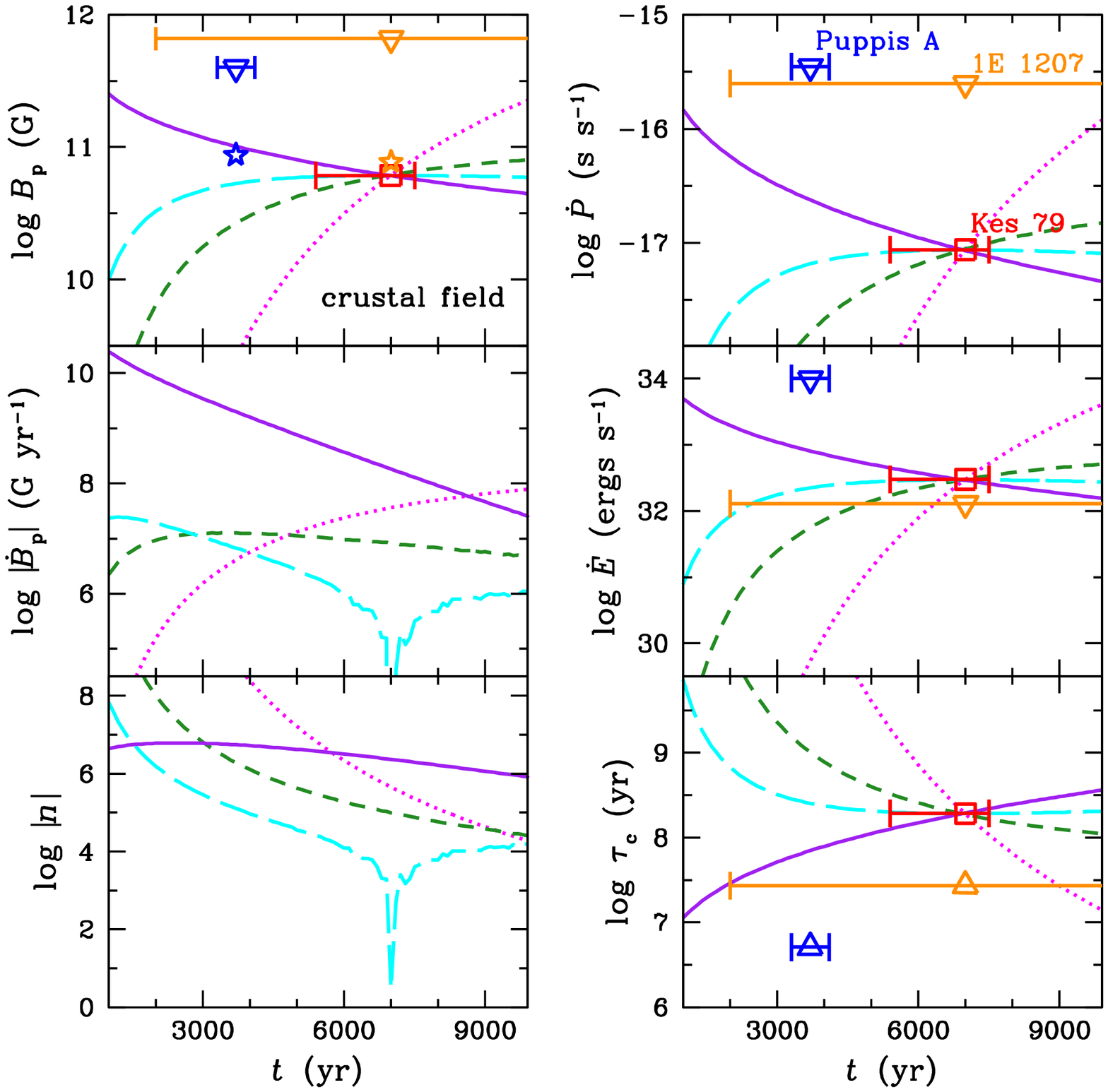}}
\caption{
Pulsar observables: magnetic field at the pole $\Bp$,
spin period derivative $\Pdot$,
rotational energy loss rate $\Edot$.
and characteristic age $\tau_\mathrm{c}$
for the CCOs Puppis~A, 1E~1207, and Kes~79;
squares are values obtained from timing measurements,
triangles are upper/lower limits obtained from timing, and
stars are from spectral measurements (see Table~\ref{tab:cco}).
Lines are evolution models with an initial spin period
$P_\mathrm{i}=0.105$~s and buried crustal field at density $\denssub$
and birth magnetic field $\Bstar$:
$[\log\denssub\mbox{(g cm$^{-3}$)},\Bstar\mbox{(G)}]=$
(8.0,$3.2\times 10^{13}$; solid),
(12.0,$3.2\times 10^{11}$; long-dashed),
(12.5,$3.7\times 10^{11}$; short-dashed), and
(13.0,$4.5\times 10^{13}$; dotted).
Also shown are the calculated magnetic field derivative $\Bdot$ and
braking index $n$; the singularity is because of a change in sign of
$\Bdot$, from a growing to decaying $\Bp$.
}
\label{fig:obs_kes79}
\end{figure}

Figures~\ref{fig:obs_pupa}-\ref{fig:obs_kes79} show the observed values and
constraints of the three CCOs from Table~\ref{tab:cco}.
The birth magnetic field $\Bstar$ is set by the normalized field
$b_r(R)$ for each submergence depth at the nominal age of the CCO
(3700~yr for Puppis~A and 7000~yr for 1E~1207 and Kes~79),
so that $\Bp=\Bstar b_r(R)$ [see eq.~(\ref{eq:magb})].
Since $\Pdot$ is directly proportional to $\Bp$ [see eq.~(\ref{eq:pdot})]
and $\Edot$ and $\tau_\mathrm{c}$ are simply related to $\Pdot$
[see eqs.~(\ref{eq:edot}) and (\ref{eq:tc})], the evolutionary tracks for
these parameters all cross at the age of the CCO.
However, we see that $\Bdot$ and $n$ do not coincide at the age of each
CCO and thus a measurement of either of these would allow one to
distinguish between the different combinations of birth magnetic
field and submergence depth.

For a purely crustal field and shallow submergence
($\denssub\lesssim3\times 10^{11}\mbox{ g cm$^{-3}$}$),
the magnetic field decreases after $t\approx 2\times 10^3$~yr
(see also Fig.~\ref{fig:evolb}).
The magnetic field decreases more rapidly the shallower the submergence,
e.g., $\Bdot\sim 10^8-10^9\mbox{G yr$^{-1}$}$
for $\denssub\approx 10^{8}\mbox{ g cm$^{-3}$}$.
On the other hand, for $\denssub\gtrsim 10^{12}\mbox{ g cm$^{-3}$}$,
the magnetic field increases more rapidly the deeper the submergence.

\begin{figure}
\resizebox{\hsize}{!}{\includegraphics{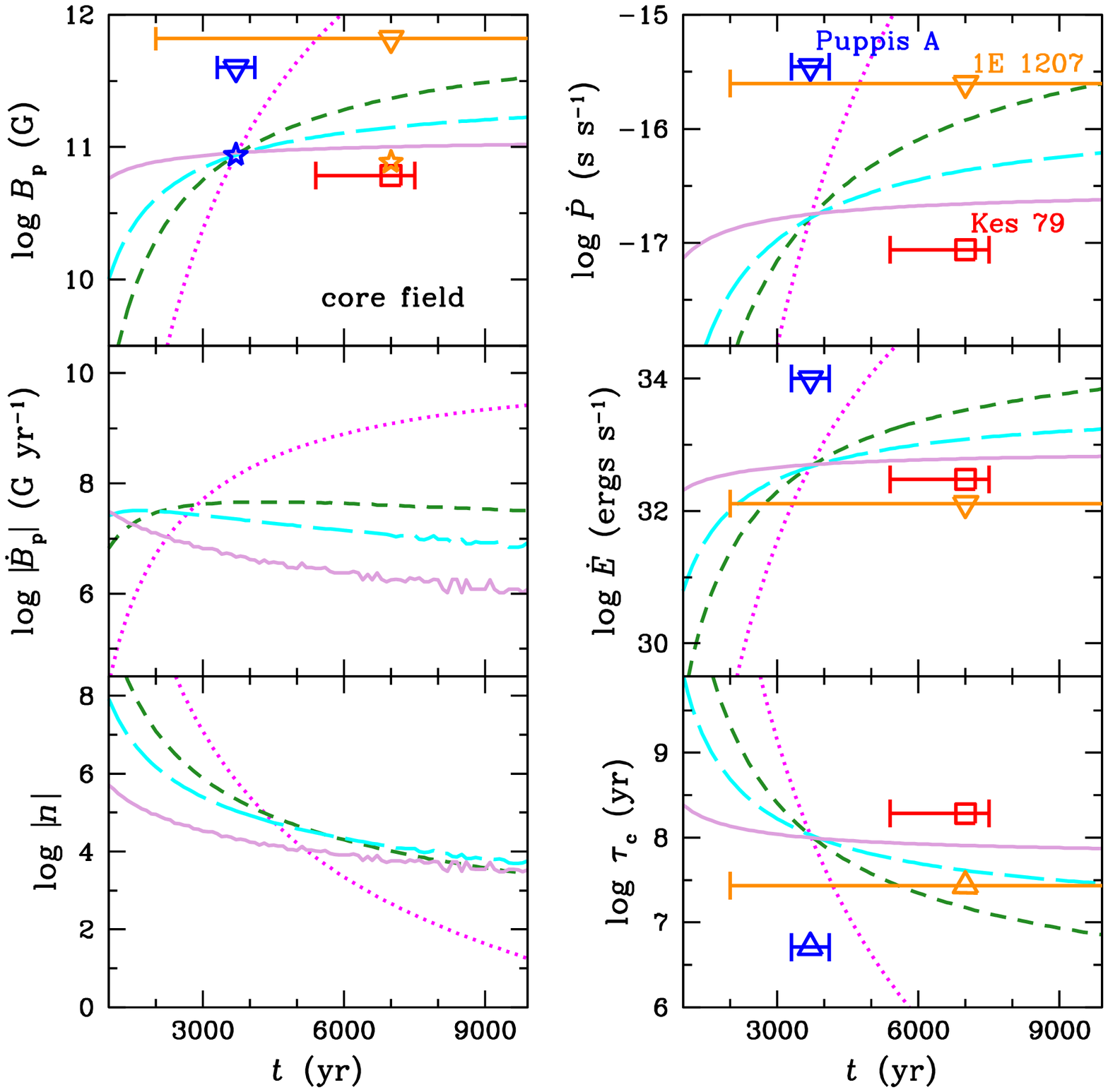}}
\caption{
Pulsar observables: magnetic field at the pole $\Bp$,
spin period derivative $\Pdot$,
rotational energy loss rate $\Edot$.
and characteristic age $\tau_\mathrm{c}$
for the CCOs Puppis~A, 1E~1207, and Kes~79;
squares are values obtained from timing measurements,
triangles are upper/lower limits obtained from timing, and
stars are from spectral measurements (see Table~\ref{tab:cco}).
Lines are evolution models with an initial spin period
$P_\mathrm{i}=0.112$~s and buried core field at density $\denssub$
and birth magnetic field $\Bstar$:
$[\log\denssub\mbox{(g cm$^{-3}$)},\Bstar\mbox{(G)}]=$
(11.0,$1.2\times 10^{11}$; solid),
(12.0,$2.8\times 10^{11}$; long-dashed),
(12.5,$1.1\times 10^{12}$; short-dashed), and
(13.0,$1.5\times 10^{15}$; dotted).
Also shown are the calculated magnetic field derivative $\Bdot$ and
braking index $n$.
}
\label{fig:obs_pupa_bc1}
\end{figure}

For a core field and deep submergence
($\denssub\gtrsim 3\times 10^{12}\mbox{ g cm$^{-3}$}$),
the surface magnetic field is still growing, and
the evolutionary tracks are almost identical to the case of the purely
crustal field.
However, for shallow submergence, the surface field grows more
quickly and saturates on the timescales considered here
($\le 10^4$~yr; see Fig.~\ref{fig:evolb}).
An example for the case of Puppis~A is shown in Fig.~\ref{fig:obs_pupa_bc1}.
Note that the field will decay on the much longer timescale set by
Ohmic decay in the core [see eq.~(\ref{eq:tohm})].

\begin{figure}
\resizebox{\hsize}{!}{\includegraphics{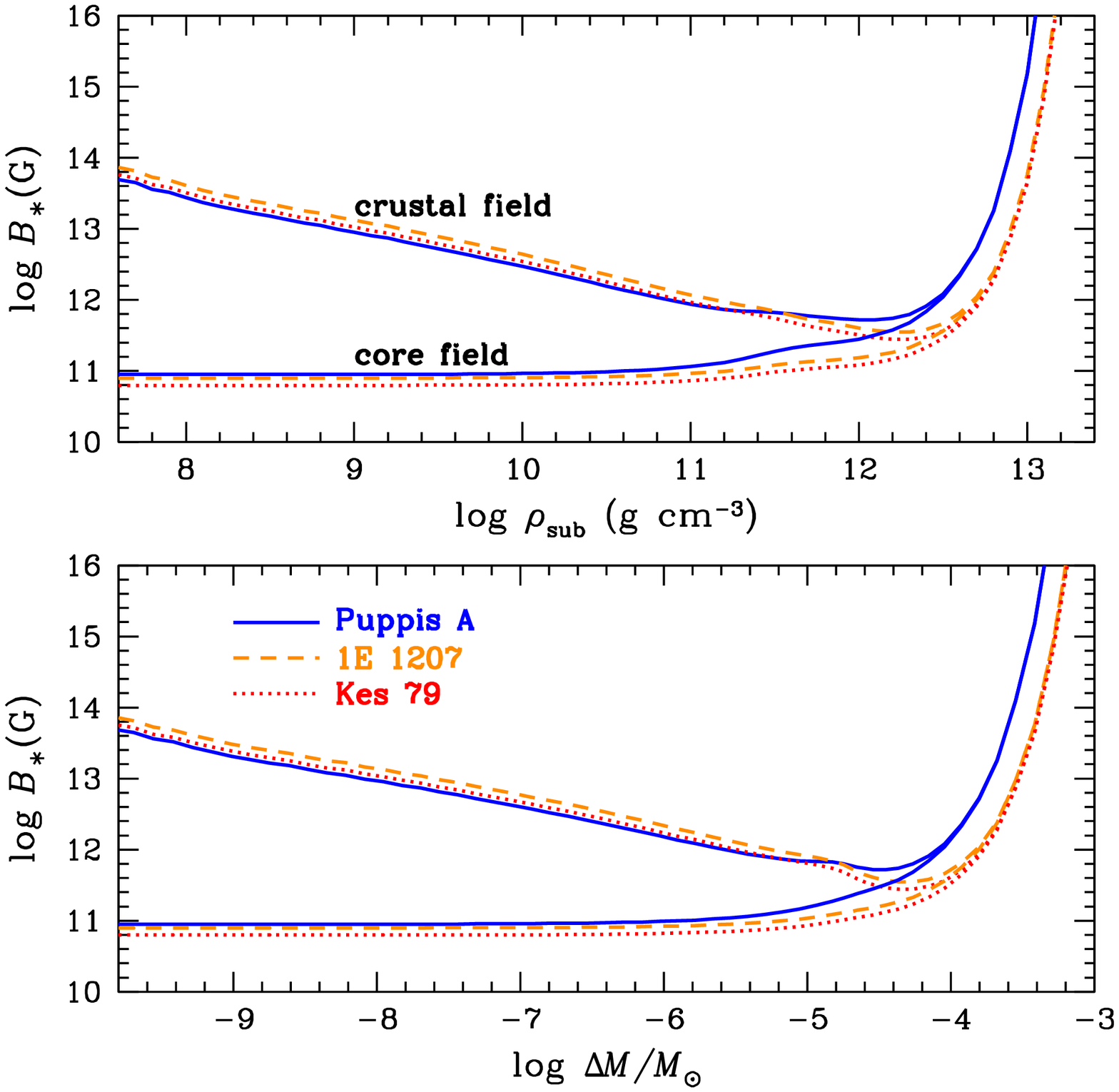}}
\caption{
Intrinsic neutron star magnetic field $\Bstar$
as a function of density at submergence $\denssub$ (top panel) and
accreted mass $\Delta M$ (bottom panel)
for the CCOs Puppis~A (solid lines), 1E~1207 (dashed lines),
and Kes~79 (dotted lines).  The upper and lower sets of lines are
for the crustal and core field configurations, respectively.
}
\label{fig:bd}
\end{figure}

Our results are summarized in Fig.~\ref{fig:bd}, where we show the inferred
birth magnetic field $\Bstar$ as a function of submergence density
$\denssub$ and accreted mass $\Delta M$ for the three CCOs, assuming
the field is either confined to the crust or determined by the core.
A measurement of both the sign and (non-zero) magnitude of $\Bdot$ can yield
a unique solution for $\Bstar$ and $\denssub$ or $\Delta M$.
A negative $\Bdot$ can only produced in a purely crustal field configuration.
A large positive $\Bdot$ is the result of deep submergence in both the
crustal and core fields, with the magnitude being the same in both cases.
There is a small density range
[$\denssub\approx (1-3)\times 10^{12}\mbox{ g cm$^{-3}$}$,
where $\Bstar\approx (1-3)\times 10^{11}\mbox{ G}$]
in the crustal field configuration where $\Bdot=0$.
For the core field, $\Bdot=0$ for $\denssub\ll 10^{11}\mbox{ g cm$^{-3}$}$,
such that the birth field is the currently measured surface field
[$\Bstar = (6-9)\times 10^{10}\mbox{ G}$].

\subsection{CCOs as an unified source} \label{sec:ccoone}

\begin{figure}
\resizebox{\hsize}{!}{\includegraphics{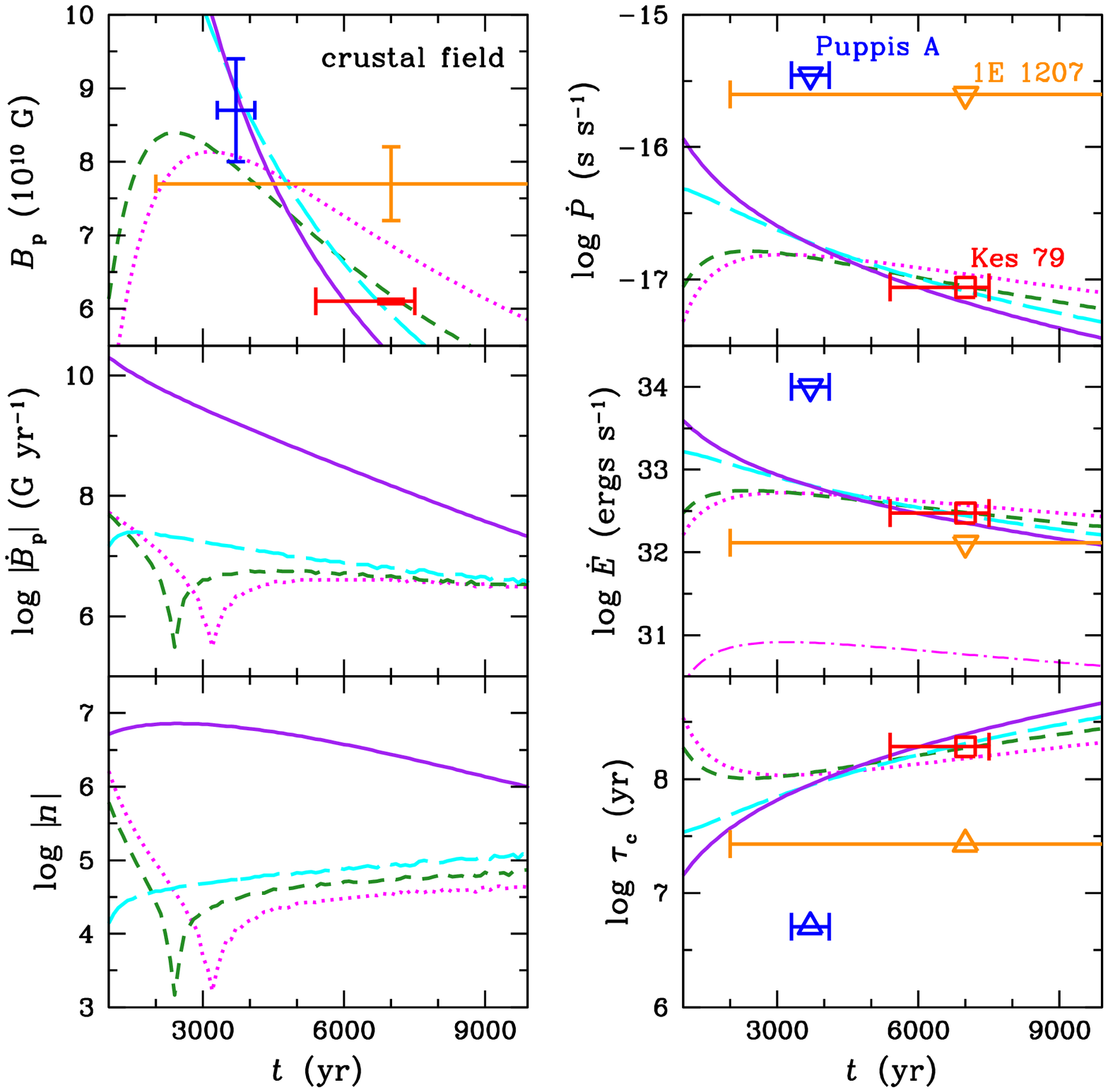}}
\caption{
Pulsar observables: magnetic field at the pole $\Bp$,
spin period derivative $\Pdot$,
rotational energy loss rate $\Edot$.
and characteristic age $\tau_\mathrm{c}$
for the CCOs Puppis~A, 1E~1207, and Kes~79;
squares are values obtained from timing measurements,
triangles are upper/lower limits obtained from timing, and
stars are from spectral measurements (see Table~\ref{tab:cco}).
Lines are evolution models with an initial spin period
$P_\mathrm{i}=0.105$~s and buried crustal field at density $\denssub$
and birth magnetic field $\Bstar$:
$[\log\denssub\mbox{(g cm$^{-3}$)},\Bstar\mbox{(G)}]=$
(7.6,$5.1\times 10^{13}$; solid),
(11.0,$9.0\times 10^{11}$; long-dashed),
(11.4,$6.2\times 10^{11}$; short-dashed), and
(11.5,$6.2\times 10^{11}$; dotted).
Also shown are the calculated magnetic field derivative $\Bdot$ and
braking index $n$; the singularities are because of a change in sign of
$\Bdot$, from a growing to decaying $\Bp$.
The dot-dashed line for $\Edot$ is the same as the dotted line but
reduced by a factor of $4^3$ due to the significantly longer spin
period of 1E~1207 [see eq.~(\ref{eq:edot})].
Note the different scale of the y-axes compared to those in
Figs.~\ref{fig:obs_pupa}-\ref{fig:obs_pupa_bc1}, especially
the linear (versus logarithmic) scale of the $\Bp$-panel.
}
\label{fig:fitone}
\end{figure}

Taken together, the three CCOs show a non-monotonic trend (based on age)
in their periods and magnetic fields.
However, there is large uncertainty in the age estimates.
Therefore let us consider the CCOs as a unified source at three different
evolutionary epochs, e.g., the age of 1E~1207 could be $\approx 5000$~yr
and the age of Kes~79 could be $\approx 6-7000$~yr.
Since field decay is required in order to match the observed surface
fields and (assumed) ages of the three CCOs, core field configurations
cannot provide a solution.
In Figure~\ref{fig:fitone}, we fit a single (crustal field) evolutionary
track to the three measured magnetic fields.
We see that for submergence densities
$\denssub\ge 3\times 10^{11}\mbox{ g cm$^{-3}$}$, the tracks
are incompatible with the fields of all three CCOs.
Since the birth magnetic field is anti-correlated with submergence
depth in the decaying field regime (see Fig.~\ref{fig:bd}), we obtain
the constraint $\Bstar\gtrsim 6\times 10^{11}$~G.
Note that the maxima at $\Bp\sim 8\times 10^{10}$~G for $\log\denssub=11.4$
and 11.5 are produced from $\Bstar=6.2\times 10^{11}$~G;
the lower surface values are the result of field decay in the NS interior
(see Sec.~\ref{sec:results}).
Finally, measuring $\Bdot$ can provide a unique solution to $\Bstar$ and
$\denssub$ or $\Delta M$ (see Sec.~\ref{sec:ccothree}).

\section{Discussion} \label{sec:discuss}

We have solved the induction equation to determine the evolution of the
internal magnetic field of a neutron star,
in order to model the field behavior (especially growth of the surface field
by diffusion) in neutron stars with ages $<10^4$~yr.
The field has a birth magnitude $\approx\Bstar$ and is initially
submerged below the neutron star surface at a density $\denssub$.
We considered both a field that is confined purely to the crust
and a field that saturates to a constant level in the core; the
latter is valid since the Ohmic decay timescale in the core is
$\gg 10^4$~yr.
Our study builds upon and improves previous works
(see, e.g., \citealt{muslimovpage95,geppertetal99}) by making use of the
latest calculations of the electrical and thermal conductivities of
matter in the density-temperature regime relevant to neutron star crusts
\citep{potekhinetal99,cassisietal07,chugunovhaensel07},
and we applied the results to interpret observations of the
recently-recognized class of central compact object neutron stars,
which have magnetic fields ($B\approx 10^{11}$~G) that are lower
than those detected in most pulsars.

For the three CCOs with measured spin period derivative $\Pdot$ or polar
magnetic field strength $\Bp$,
we showed that there is a well-defined relationship between the
birth magnetic field $\Bstar$ and submergence density $\denssub$
(or accreted mass $\Delta M$) in the case of a purely crustal field;
a well-defined relationship also exists for the case of a core field
but only at $\denssub\gtrsim 10^{11}\mbox{ g cm$^{-3}$}$.
Population synthesis analyses indicate a normal distribution for birth
spin periods $P_\mathrm{i}$ and a lognormal distribution for birth
magnetic fields:
If no long-term field decay occurs, then the peak and width of the
distributions are $P_\mathrm{i}=300$~ms and $\sigma=150$~ms and
$\log\Bstar\mbox{(G)}=12.95$ with $\sigma=0.55$ \citep{fauchergiguerekaspi06},
while simulations including field decay result in
$P_\mathrm{i}=250$~ms and $\sigma=100$~ms
and $\log\Bstar=13.25$ with $\sigma=0.6$ \citep{popovetal10}.
If the CCOs are born with birth magnetic fields that follow
these distributions (e.g., $\log\Bstar\gtrsim 12$ at 2$\sigma$),
then the field is buried at large depths
($\denssub\gtrsim 2\times 10^{12}\mbox{ g cm$^{-3}$}$
or $\Delta M\gtrsim 10^{-4}M_\odot$) for both the crustal and core field
configurations or at shallow depths in the crustal field configuration
($\denssub<10^{11}\mbox{ g cm$^{-3}$}$ or $\Delta M\lesssim 10^{-5}M_\odot$).
Incidentally, \citet{gotthelfhalpern07} argue that the slow spin and weak
magnetic field of 1E~1207 make it possible for accretion from a debris disc left
over after the supernova.
However, recent optical and infrared observations of 1E~1207
place strong limits on this disc, including an (model-dependent)
estimate of the initial and current total disc mass of
$\lesssim 10^{-6}\,M_\odot$ and $\lesssim 10^{-10}\,M_\odot$, respectively
\citep{delucaetal11},
which implies that the magnetic field could only be buried to a maximum
density of $\sim 10^{11}\mbox{ g cm$^{-3}$}$.

We showed that different combinations of birth magnetic field
and submergence depth can lead to the same observed values of the spin
period derivative, as well as surface magnetic field, rate of rotational
energy loss, and characteristic age.
Predicted values of the braking index or rate of change of the magnetic
field $\Bdot$ could distinguish between these combinations.
Although the braking index is not likely to be detectable in the CCOs,
it may be possible (though difficult) to measure the rate of change of
the magnetic field $\Bdot$ from the spectral lines
(in the case of Kes~79, where the magnetic field has only been determined
from a $\Pdot$ measurement, an electron cyclotron line would occur at
$\sim 0.5-0.6$~keV).
Only a crustal field configuration can produce a negative $\Bdot$,
while a large positive $\Bdot$ indicates deep submergence, irrespective of
field configuration.
If $\Bdot$ is constrained to be very small, then
the submergence density is either shallow for a core field or
$\approx 10^{12}\mbox{ g cm$^{-3}$}$
($\Delta M\approx 5\times 10^{-5}M_\odot$) for a crustal field.

If the three CCOs are treated as a single source at different epochs,
then the surface field is seen to decay with time, which rules out a
core field configuration since the Ohmic decay timescale is much longer
than $10^4$~yr.
For a purely crustal field, we showed that
$\denssub\lesssim 3\times 10^{11}\mbox{ g cm$^{-3}$}$ and
$\Bstar\gtrsim 6\times 10^{11}$~G.
A constraint at the other end can be obtained by considering
another member of the CCOs, the neutron star in the Cassiopeia~A
supernova remnant, which has an age of $330\pm 19$~yr \citep{fesenetal06}.
The field of the Cas~A CCO was found to be $<10^{11}$~G from fits to
its X-ray spectrum \citep{hoheinke09,heinkeho10,shterninetal11};
the non-detection of pulsations from this source
\citep{murrayetal02,mereghettietal02,ransom02,pavlovluna09,halperngotthelf10}
also indicates that the field is too low to produce pulsar-like emission.
Its low field at a young age suggests $\denssub>10^{11}\mbox{ g cm$^{-3}$}$
and $\Bstar<9\times 10^{11}$~G.
A better determination of the ages would (in)validate the CCOs as
a unified source and could allow for a stronger constraint on their
birth magnetic fields.

We also note that, irrespective of our magnetic field evolution calculations,
the $\Pdot$ that is inferred from the spectral lines of Puppis~A and 1E~1207
result in $\Edot\approx 4.4\times 10^{32}$ and
$1.7\times 10^{30}\mbox{ ergs s$^{-1}$}$, respectively, while
$\Edot=3.0\times 10^{32}\mbox{ ergs s$^{-1}$}$ from the timing of Kes~79.
Standard neutron star cooling (see Sec~\ref{sec:evolt}) yields
the redshifted bolometric luminosity $L^\infty_\mathrm{bol}$, so that
$\Edot/L_\mathrm{bol}^\infty\sim 0.2-0.3$ for Puppis~A and Kes~79 and a
much lower value of $\sim 10^{-3}$ for 1E~1207 (due to its slower spin period).
Furthermore, $\Edot^{1/2}/d^2\sim 10^{-4}$ (relative to Vela) for
Puppis~A and $10^{-5}$ for 1E~1207 and Kes~79, where $d$ is the source
distance;
these are far below the values from sources that have been detected
in the gamma-rays by {\it Fermi}
(\citealt{smithetal08,abdoetal10}; see also \citealt{zaneetal11}).

Finally, \citet{halperngotthelf10,kaspi10} noted that CCOs occupy an
underpopulated region in $P-\Pdot$.
On the one hand, we have shown that the CCOs may just be representative
of the low end of the distribution of birth magnetic fields,
with $B\sim 10^{12}$~G.
On the other hand, CCOs may have higher magnetic fields that have been
submerged to great depths.
In this case, $\Pdot$ is increasing rapidly, and the CCOs are evolving
to join the majority of the pulsar population at longer spin periods,
higher $\Pdot$, and higher observed magnetic fields.

\section*{acknowledgements}

WCGH thanks the referee, Ulrich Geppert, for comments that improved the
clarity of the manuscript.
WCGH appreciates the use of the computer facilities at the Kavli
Institute for Particle Astrophysics and Cosmology.
WCGH acknowledges support from the Science and Technology Facilities
Council (STFC) in the United Kingdom.

\bibliographystyle{mnras}

\label{lastpage}

\end{document}